\documentclass[9pt,twocolumn,twoside]{pnas-new}
\usepackage{color}

\templatetype{pnasresearcharticle} 
\providecommand{\e}[1]{\ensuremath{\times 10^{#1}}}
\title{Auxetic metamaterials from disordered networks}

\author[a]{Daniel R Reid}
\author[b]{Nidhi Pashine}
\author[c]{Justin M Wozniak}
\author[b]{Heinrich M Jaeger}
\author[d]{Andrea J Liu}
\author[b]{Sidney R Nagel}
\author[a]{Juan J de Pablo}

\affil[a]{Institute for Molecular Engineering, University of Chicago}
\affil[b]{James Frank Institute, University of Chicago}
\affil[c]{Argonne National Laboratory}
\affil[d]{Department of Physics, University of Pennsylvania}

\leadauthor{Reid}

\authorcontributions{D.R. Reid and J.J. de Pablo designed the model and calculations. D.R. Reid carried out the simulations.  N. Pashine created and characterized experimental networks.  J.M. Wozniak assisted in parallelizing simulation workflow.  H.M. Jaeger, A.J. Liu, S.R. Nagel and J.J. de Pablo planned and conceived the study.}
\authordeclaration{The authors declare no conflicts of interest.}

\correspondingauthor{\textsuperscript{5}To whom correspondence should be addressed. E-mail:depablo@uchicago.edu}

\begin{abstract}
Recent theoretical work suggests that systematic pruning of disordered networks consisting of nodes connected by springs can lead to materials that exhibit a host of unusual mechanical properties.  In particular, global properties such as the Poisson's ratio or local responses related to deformation can be precisely altered. Tunable mechanical responses would be useful in areas ranging from impact mitigation to robotics and, more generally, for creation of metamaterials with engineered properties. However, experimental attempts to create auxetic materials based on pruning-based theoretical ideas have not been successful.  Here we introduce a new and more realistic model of the networks, which incorporates  angle-bending forces and the appropriate experimental boundary conditions. A sequential pruning strategy of select bonds in this model is then devised and implemented that enables engineering of specific mechanical behaviors upon deformation, both in the linear and non-linear regimes. In particular, it is shown that the Poisson's ratio can be tuned to arbitrary values. The model and concepts discussed here are validated by preparing physical realizations of the networks designed in this manner, which are produced by laser cutting two-dimensional sheets and are found to behave as predicted. Furthermore, by relying on optimization algorithms, we exploit the networks' susceptibility to tuning to design networks that posses a distribution of stiffer and more compliant bonds, and whose auxetic behavior is even greater than that of homogeneous networks.  Taken together, the findings reported here serve to establish that pruned networks represent a promising platform for the creation of novel mechanical metamaterials.
\end{abstract}

\begin{document}

\verticaladjustment{-2pt}

\maketitle
\thispagestyle{firststyle}
\ifthenelse{\boolean{shortarticle}}{\ifthenelse{\boolean{singlecolumn}}{\abscontentformatted}{\abscontent}}{}

\dropcap{W}hen one stretches a material along one axis, intuition suggests that the material will contract in the orthogonal lateral directions.  For most natural and synthetic materials, this intuition is confirmed by experiment.  This behavior is quantified by the Poisson's ratio, $\nu$, which for a deformed material is defined as the negative ratio of the material's lateral strain to its axial strain.   In linear elastic theory for an isotropic sample, the Poisson's ratio is a monotonic function of the ratio of the material's shear modulus, $G$, to its bulk modulus, $B$.  In two dimensions $\nu \rightarrow 1$ as $G/B \rightarrow 0$.  In this limit, the material is `incompressible', meaning that its volume does not change during this axial compression.  In the limit $G/B \rightarrow \infty$,  $\nu \rightarrow -1$.  In the range where $\nu$ is negative, materials become wider as they are stretched, and thinner as they are compressed.  Such materials, termed `auxetics', show promise in applications that require structural elements\cite{evans1991design, miller2011honeycomb, evans2004design}, impact absorbers \cite{alderson1994auxetic, sanami2014auxetic}, filters\cite{alderson2000auxetic, alderson2001auxetic}, fabrics\cite{alderson2012auxetic, hu2011development}, or other, tailor-made mechanical responses.
\paragraph{}

Auxetic materials have been formed through a variety of preparation protocols.  Under special processing conditions, polymer foams and fibers, for example, can exhibit negative Poisson's ratios\cite{lakes1987foam, caddock1989microporous, chan1997fabrication, ravirala2006negative, alderson1994auxetic}.
Auxetic foams, in particular, can be formed through a process of heating and sintering fine particles of ultra-high molecular weight polyethylene \cite{lakes1987foam, pickles1995effect}, leading to structures of nodes connected by thin fibrils which collapse isotropically when compressed.  Such structures are termed ``re-entrant'', and are a common motif in auxetic materials \cite{lakes1987foam, chan1997microscopic}. When compressed uniaxially, these nodes and fibrils undergo complex rearrangements that give rise to their auxetic behavior. As materials approach the lower limit of the Poisson's ratio, their hardness, or resistance to a small indentation, is predicted to increase rapidly\cite{timoshenkotheory}.  This prediction is confirmed in the case of ultra-high molecular weight polyethylene, where the hardness of the auxetic material far exceeds that of a non-auxetic but otherwise equivalent foam \cite {alderson1994auxetic}.
\paragraph{}

The node and fibril structures common in auxetic polymer foams can be thought of as networks consisting of nodes connected by bonds.  A central, common feature of past efforts to design auxetic materials in both theory and experiment, however, has been a reliance on regular, ordered lattices.
Such lattices include the double arrowhead structure\cite{larsen1997design, alderson2012auxetic}, star honeycomb structures \cite{theocaris1997negative}, re-entrant honeycombs\cite{gibson1982mechanics, warren1990negative}, and others\cite{florijn2014programmable}.  Building on recent theoretical arguments\cite{goodrich2015principle,hexner2017role,hexner2017linking}, in this work we focus on disordered, random networks.
\paragraph{}

In the linear regime, the bulk modulus, $B$, or the shear modulus, $G$, of a network are proportional to the sum of the potential energies that are stored in each bond when the network is compressed or sheared.  The decrease in $B$ or $G$ when the $i^{th}$ bond is removed is denoted $\Delta B_{i}$ or $\Delta G_{i}$, respectively.  In a simple crystalline network, every bond responds in nearly the same manner to a global deformation.  In contrast, in amorphous networks the response of individual bonds to a global deformation can span many orders of magnitude \cite{goodrich2015principle,hexner2017linking}.

Furthermore, there is little correlation between the value of $\Delta B_{i}$ and $\Delta G_{i}$ of a bond, $i$ \cite{hexner2017linking}.  This suggests that, by selective removal or ``pruning'' of bonds with large or small values of $\Delta G_i$ or $\Delta B_i$, the ratio $G/B$ can be manipulated to reach a desired value; this would lead to disordered, ``amorphous'' materials with intriguing mechanical properties.  Recent work has shown that similar pruning strategies could be used to design allosteric interactions into a network (where a  deformation at a local source can produce a desired response at a distant target site).  This behavior was demonstrated in experiments \cite{rocks2017designing}. Creating auxetic materials, however, is more challenging and success in creating experimental prototypes has been elusive.  More specifically, simple models were used to design pruned networks with negative Poisson's ratio but, when prepared in the laboratory, they failed to exhibit auxetic behavior. This state of affairs has led to the question of whether pruning based approaches for design of auxetic materials are fundamentally flawed, or whether it is indeed possible to engineer truly auxetic laboratory materials by relying on more sophisticated models. 

\paragraph{}

Here we address that question by introducing a mechanical model of disordered networks that incorporates the effects of angle-bending in a novel way.  The model is minimally complex, and it is parameterized by comparison to experimental data for simple, random disordered networks. By adopting a pruning strategy that identifies and removes select bonds from these networks, it is shown that it is possible to reach Poisson's ratios as low as $\nu=-0.8$. The two-dimensional pruned networks designed in this manner are then prepared in the laboratory from rubber sheets that have been laser-cut according to the simulated models. They are found to behave as predicted. Structural analysis shows that highly auxetic networks are marked by an abundance of concave polygons.  When networks are compressed uniaxially, these concave polygons shrink in all dimensions.  Collectively, the local deformations of these concave polygons yield global auxetic behavior.  These structures also give rise to a sub-linear stress-strain behavior, which is an important characteristic of impact-mitigating materials. We also investigate the effect of bond-bending stiffness - a critical parameter used to inform the construction of experimental networks - on a materials' ability to be made auxetic.  Such a parameter is also relevant in the creation of auxetic foams.  We find that networks with bonds of extremely low bending stiffness can be tuned to show a Poisson's ratio near $\nu=-1.0$, while networks with much stiffer bond-bending forces cannot be tuned at all.  Such changes are explained by the distributions and correlations of $\Delta G_{i}$ and $\Delta B_{i}$. We conclude our discussion by designing highly auxetic materials through a materials optimization strategy. Specifically, by selectively manipulating the mechanical characteristics of a few select bonds, it is shown that values of the Poisson's ratio as low as $\nu=-0.9$ can be achieved. The improved networks designed in this manner can then be successfully produced in the laboratory.

\section*{Models}

\subsection*{Simulation Model}

Networks are created from disordered jammed packings of frictionless spheres at zero temperature created using standard procedures \cite{liu2010jamming}.  Spherical particles are initially placed at random positions within the simulation area.  Particles $i$ and $j$ experience harmonic repulsions:

\begin{equation}
V(r_{ij}) = \frac{\epsilon}{2} (1-\frac{r_{ij}}{\sigma_{ij}})^{2} \Theta\left(1-\frac{r_{ij}}{\sigma_{ij}}\right)
\label{eqn:softSphere}
\end{equation}
where $r_{ij}$ is the center-to-center distance, $\sigma_{ij}$ is the sum of the radii of particles $i$ and $j$, and $\Theta(x)$ is the Heaviside step function.  $\epsilon=1$ sets the energy scale.  The energy is minimized to produce zero-temperature, mechanically stable configurations.
Particles are randomly assigned one of four evenly spaced radii (namely 0.6, 0.74, 0.87 and 1.0), leading to an amorphous packing when compressed isotropically.  In all calculations, the contacts of particles which are in contact with fewer than three adjacent particles are not counted towards the total $Z$, as these would not contribute to the modulus of a jammed system. Such particles are removed before bonds are formed.

\begin{figure}
	\centering
	\includegraphics[width=4cm]{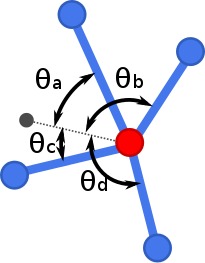}
	\caption{Schematic describing how angular restraints are applied for the node shown in red.  A director, shown in grey, is attached to each node with a harmonic bond potential.  Harmonic angles potentials are added between each pair of bonded nodes and the director, as indicated by the angles $\theta_{a-d}$. The director is positioned such that $\theta_{a-d}$ are as far from $0^{\circ}$ and $180^{\circ}$ as possible.  This scheme is applied at each node.}

	\label{fig:director}
\end{figure}

\paragraph{}

Two particles are considered to be in contact when $r_{ij}<\sigma_{ij}$. The average number of contacts or bonds per particle, $Z$, plays a central role in a host of network characteristics \cite{liu2010jamming,sussman2016spatial,driscoll2016role}.  To set the value of $Z$, we create harmonic repulsive walls at the simulation box edges, whose positions are adjusted and the configuration relaxed until the required number of particle contacts is achieved.  Unstretched bonds of length $r_{ij}^{0}$ are then placed between the centers of pairs of contacting particles $i$ and $j$ and the soft-sphere potential is removed.  The energy due to bond compression is thus:

\begin{equation}
V_{c}(r_{ij}) = \frac{1}{2r_{ij}^{0}} (r_{ij}-r_{ij}^{0})^{2} \;.
\label{eqn:compression}
\end{equation}

\paragraph{}
In order to include angle-bending constraints, we introduce a unit vector, $\vec s_i$, at each node $i$ of the network, as shown in Fig. \ref{fig:director}.  A bond connecting nodes $i$ and $j$, makes an angle $\theta_{ij\vec s_i}$ with the vector $\vec s_i$.  When the system is relaxed, this angle adopts its equilibrium value, $\theta_{ij\vec s_i}^{0}$. The energy cost to change an angle is quadratic:

\begin{equation}
V_b(\theta_{ij\vec s_i}) = \frac{k_{ang}}{2} (\theta _{ij\vec s_i}-\theta_{ij\vec s_i}^{0}) ^{2}
\label{eqn:anglepotential}
\end{equation}
where $k_{ang}$ sets the energy scale for the angle-bending potential.  During energy minimization, to obtain the ground state where the system is in mechanical equilibrium, the direction of $\vec s$ on each site is allowed to vary in order to minimize the total angular energy of a node.  The coefficient $k_{ang}$ is determined by comparing the response of model networks to those prepared in experiments and depends on the material and shape of the bonds, as described in Methods.

\paragraph{}
The total energy of a network under stress is the sum of two terms: a compressive component given by Eq.~\ref{eqn:compression} and a bending component, given by Eq.~\ref{eqn:anglepotential}.  Note that the compressive strength is scaled by $1/r_{ij}^{0}$ as would occur in a physical mechanical strut of constant thickness.

\begin{figure}
	\centering
	\includegraphics[width=8cm]{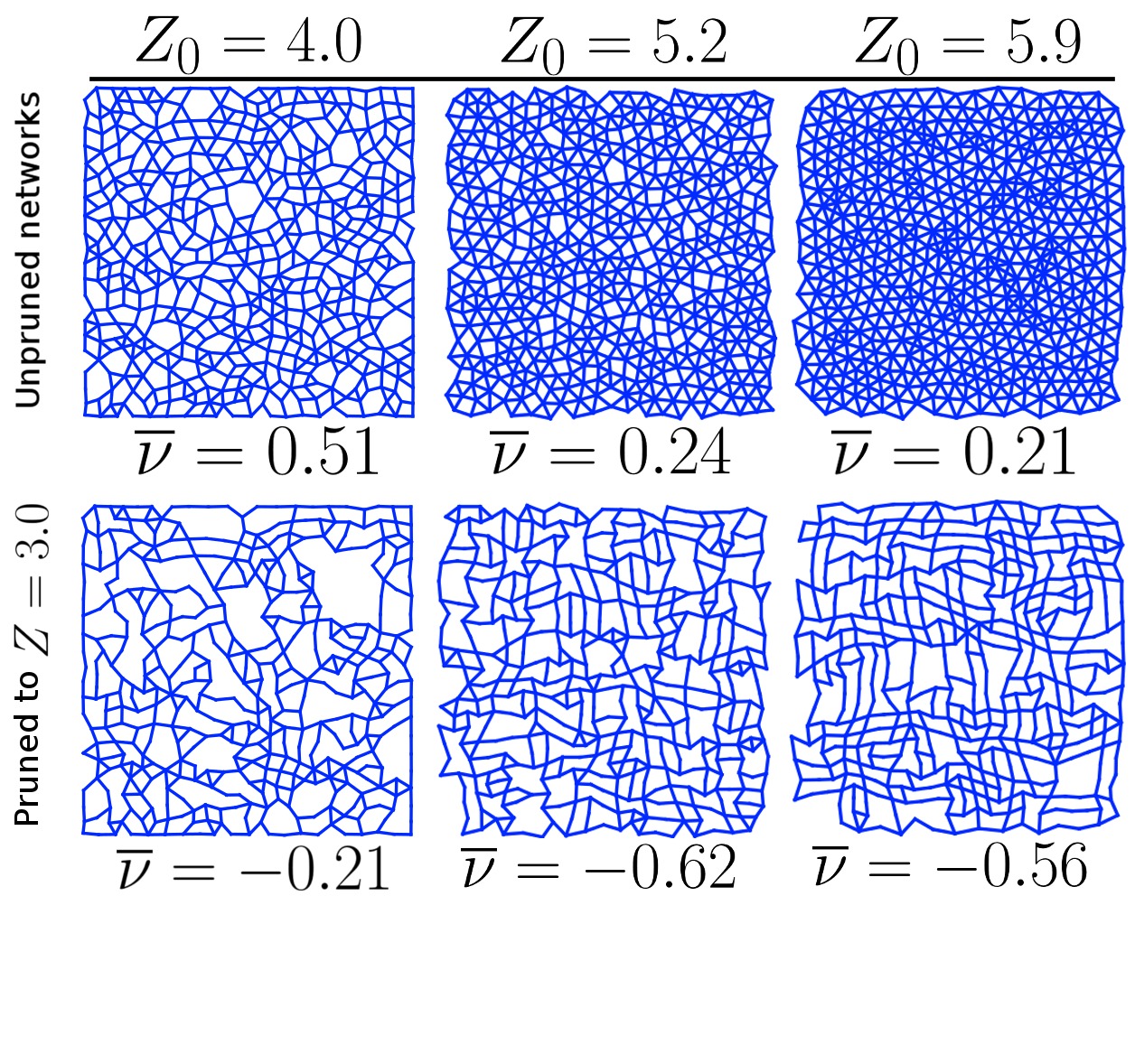}
\caption{Examples of typical 500-node networks before and after pruning with $Z_{0}=4.0, 5.2, 5.9$.  The top row shows unpruned networks, while the bottom row shows networks which have been pruned to $Z=3.0$.  Unpruned networks show decreasing $\nu$ as $Z_{0}$ increases while pruned networks show a minimum $\nu$ at $Z_{0}=5.2$.  This minimum $\nu$ at $Z_{0}=5.2$ corresponds to a high fraction of re-entrant nodes which can collapse inwards as the system is compressed. }
	\label{fig:combo}
\end{figure}

Figure \ref{fig:combo} shows representative realizations of two-dimensional disordered networks consisting of nodes connected by bonds, before and after pruning.

\paragraph{}
In two dimensions, there are two independent shear moduli - one associated with simple shear and one with pure shear.  The modulus associated with simple shear influences the value of $\nu$ that is measured when the material is deformed by pulling or pushing from opposite corners. The modulus associated with pure shear relates to the value of $\nu$ measured when the material is uniaxially compressed or expanded in $x$ or $y$, as shown in Supplementary Information.  In this study we focus primarily on algorithms that only influence the modulus associated with pure shear since this can be more easily measured in our experiments.  However we also show that isotropic auxetic networks can be created using similar algorithms as discussed in detail in the Supplementary Information.  Such materials are auxetic with respect to any uniaxial deformation. $G$ and $B$ are measured as described in Methods.
\paragraph{}

\section*{Results}
\subsection*{Bond Response Distributions}
In an amorphous network the distributions of $\Delta B_{i}$ and $\Delta G_{i}$, $P(\Delta B_{i})$ and $P(\Delta G_{i})$ can span many orders of magnitude.  That is, when some bonds are removed, $G$ or $B$ may decrease significantly, while when others are removed, there may only be a negligible decrease.  Our pruning procedure targets bonds that contribute little to the shear modulus but contributing strongly to the bulk modulus. It is therefore important that $P(\Delta B_{i})$ and $P(\Delta G_{i})$ be broad and extend to small values \cite{goodrich2015principle, hexner2017linking}.

A second crucial condition for successful pruning is that $\Delta B_{i}$ and $\Delta G_{i}$ be uncorrelated.  Based on these two features, one can selectively remove bonds from a disordered network in order to drive $B$, $G$, and thus $\nu$, to a desired target value \cite{goodrich2015principle, hexner2017role}.

\begin{figure}
	\centering
	\includegraphics[width=8cm]{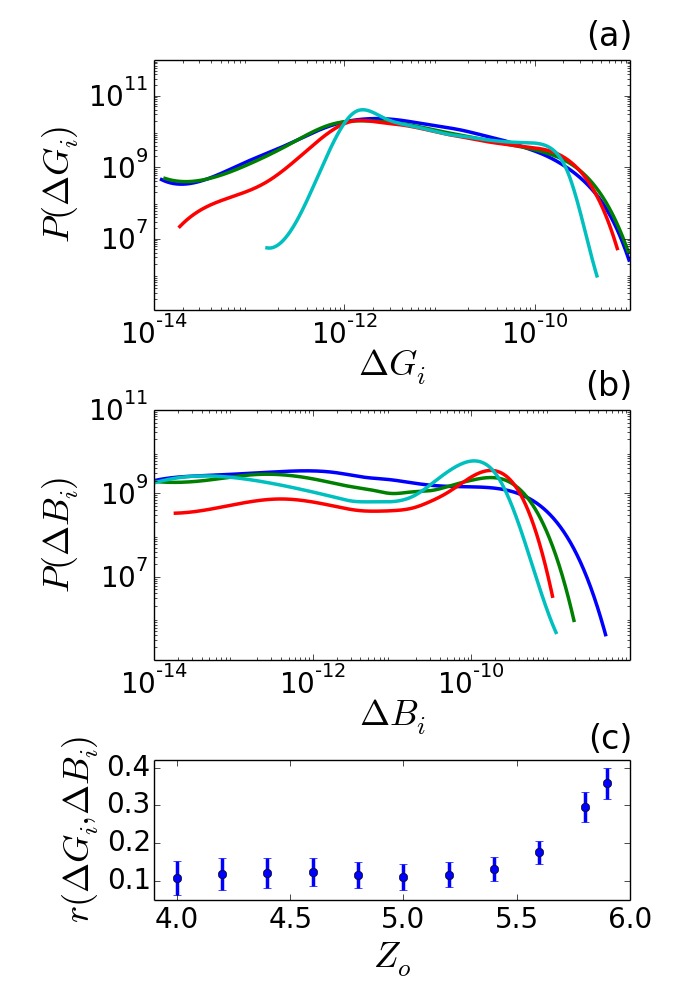}
\caption{Probability distributions and correlations of $\Delta G_{i}$ and $\Delta B_{i}$ for unpruned 500-node networks with different $Z_{0}$. Panel (a) shows probability distributions of $\Delta G_{i}$, while Panel (b) shows those for $\Delta B_{i}$.   Strains for both deformations are $\epsilon_{y}=1\e{-4}$.   Data sets as ($Z$, color) are (4.0, blue), (4.8, green), (5.2, red), and (5.9, cyan).  Each data set is taken from $100$ independent $500$ node networks.  The bond bending strength is $k_{ang} = 0.01$, a value which is experimentally realizable.  As the coordination number increases, distributions narrow significantly, reducing networks' propensity to be pruned.  Panel (c) shows the correlation between $\Delta G_{i}$ and $\Delta B_{i}$ for networks over a range of initial $Z$ values.  Standard deviation of $r$ values across $100$ independent configurations are shown.}
	\label{fig:stressDists}
\end{figure}

\paragraph{}
Panels a) and b) of Fig. \ref{fig:stressDists} show the probability distributions $P(\Delta B_{i})$ and $P(\Delta G_{i})$ for unpruned networks. Results are shown for networks with $Z_{0}$ ($Z$ of the network before pruning) between $4.0$ and $5.9$, with $k_{ang} = 0.01$.  The value of $k_{ang}$ is set as the value which best reproduces the deformation observed in experiment, as described in Methods.  As $Z_{0}$ increases, both $P(\Delta B_{i})$ and $P(\Delta G_{i})$ become narrower.  This  suggests that networks with lower coordination numbers are more amenable to pruning.  A peak in $P(\Delta B_{i})$ becomes apparent for $Z_{0}=5.9$.  

\paragraph{}

In order to facilitate effective pruning, bond response distributions must not only be broad, but uncorrelated.  Panel (c) of Fig. \ref{fig:stressDists} shows the Pearson correlation coefficient for $\Delta G_{i}$ and $\Delta B_{i}$ across a range of $Z_{0}$ values.  While distributions are significantly uncorrelated between $Z_{0}=4.0$ and $5.2$, the level of correlation increases thereafter.  As we will see, networks pruned from $Z_{0}=5.2$ lead to the lowest value of $\nu$.

\subsection*{Pruning}
For the iterative pruning strategy adopted here, at each iteration the lowest $\Delta G_{i}$ bond is removed.  Given the low correlation between $\Delta G_{i}$ and $\Delta B_{i}$, pruning the lowest $\Delta G_{i}$ bonds tends to increase $G/B$ and decrease $\nu$.  At each iteration, $\Delta G$ for a bond is measured by performing a trial removal of that bond, and by measuring the resulting shear modulus, thereby resulting in $n_{bond}$ measurements of $G$.  Each measurement of $G$ requires on the order of seconds or minutes of CPU time, depending on network size - creating a large auxetic network can therefore be computationally demanding.   The calculation of $\Delta G_{i}$ can be parallelized with one core per $\Delta G_{i}$ measurement.  We use the parallel workflow management software Swift/T to parallelize this process across several hundred CPU cores, thereby accelerating network creation \cite{wozniak2013swift}.

\begin{figure}
	\centering
	\includegraphics[width=8cm]{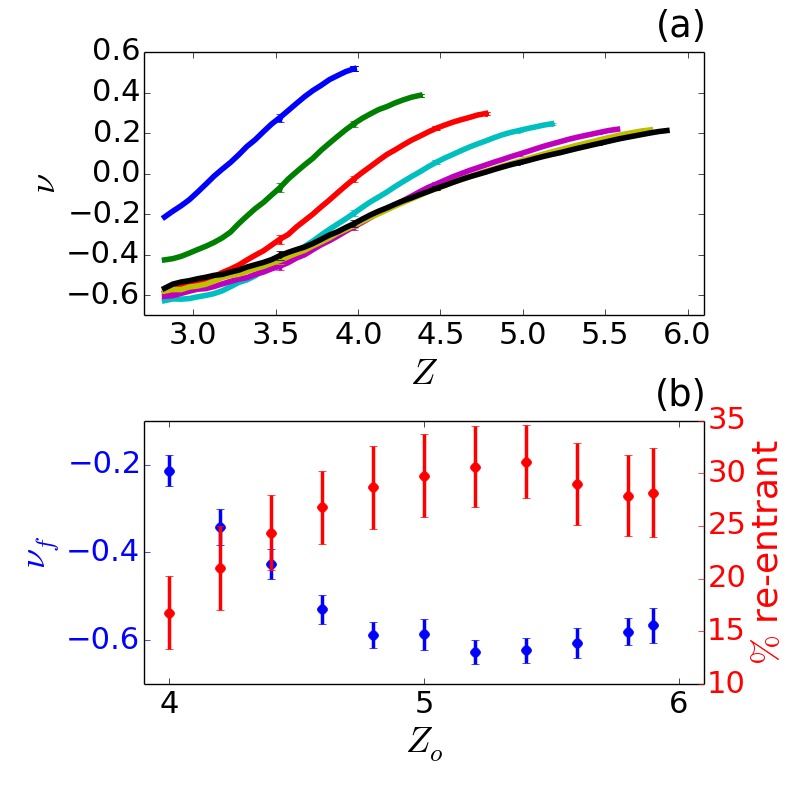}
	\caption{Poisson's ratios resulting from pruning 500-node networks with different values of $Z_{0}$ shown in (a) with resulting structural and mechanical properties shown in (b). Data are taken from $50$ independent $500$ node networks. In (a), ninety-five percent confidence intervals are shown. To prune, we remove the lowest $\Delta G$ bond at each iteration.  Networks are pruned until $Z=2.8$, at which point they become so sparse that $\nu$ fluctuates wildly with further pruning. In (b), the blue data set shows the Poisson's ratio reached at $Z=3.0$ with respect to $Z_{0}$, with $95\%$ confidence intervals shown.  A minimum $\nu$ is observed at $Z_{0}=5.2$.  Plotted in red is the fraction of nodes which are classified as re-entrant.  The data shows that the most auxetic networks show the greatest degree of re-entrant behavior, suggesting a structural origin to $\nu$.}
	\label{fig:poissons_with_prune_combo}
\end{figure}

\paragraph{}
Figure \ref{fig:poissons_with_prune_combo} shows Poisson's ratios, $\nu$, for networks having different values of $Z_{0}$.  Each data set represents an average of $50$ independent pruned networks of $500$ nodes each.

The Poisson's ratio is determined by introducing a small strain of magnitude  $\epsilon_{y}=1\e{-4}$ in the $y$ dimension of the network, allowing the system to relax to an average force tolerance of $1\e{-13}$, and then measuring the resulting lateral deformation. 
Note that while simple bead-spring networks (those ignoring angle bending) lose rigidity below $Z=4$, the angular restraints in the model introduced here lead to rigid networks to a much lower values of $Z$ .  As discussed below, $\epsilon_{y}=1\e{-4}$ is well within the linear regime.

\paragraph{}
Several interesting features are apparent in the pruning progression shown in Fig. \ref{fig:poissons_with_prune_combo}. First, even before there is any pruning, the Poisson's ratio of the networks  decreases from $0.51$ at $Z_{0}=4.0$ to $0.21$ at $Z_{0}=5.9$, revealing a wide variation of $\Delta \nu=0.3$. Second, the initial slope of the Poisson's ratio curve as a function of pruning decreases from $d\nu/dZ = 0.47$ to $0.14$ between $Z_0=4.0$ and $Z_0=5.9$ (as calculated by the average slope over the first $\Delta Z = 0.1$ pruning).  The smaller value of $d\nu/dZ$ at higher $Z_{0}$ is consistent with the narrower distribution functions and higher correlations observed, as shown in Fig. \ref{fig:stressDists}.  However, a higher $Z_{0}$ also implies that there are simply more bonds available for pruning. These factors conspire to produce the lowest pruned networks when $Z_{0}=5.2$ . This also corresponds to the highest $Z_{0}$ before the correlation of $\Delta G_{i}$ and $\Delta B_{i}$ begins to increase, as seen in Fig. \ref{fig:stressDists}c. Networks with $Z_{0}=5.2$  show a minimum average of $\nu=-0.62$. The lowest $\nu$ value achieved for an individual network, however, is $\nu=-0.79$.

\begin{figure}
	\centering
	\includegraphics[width=8cm]{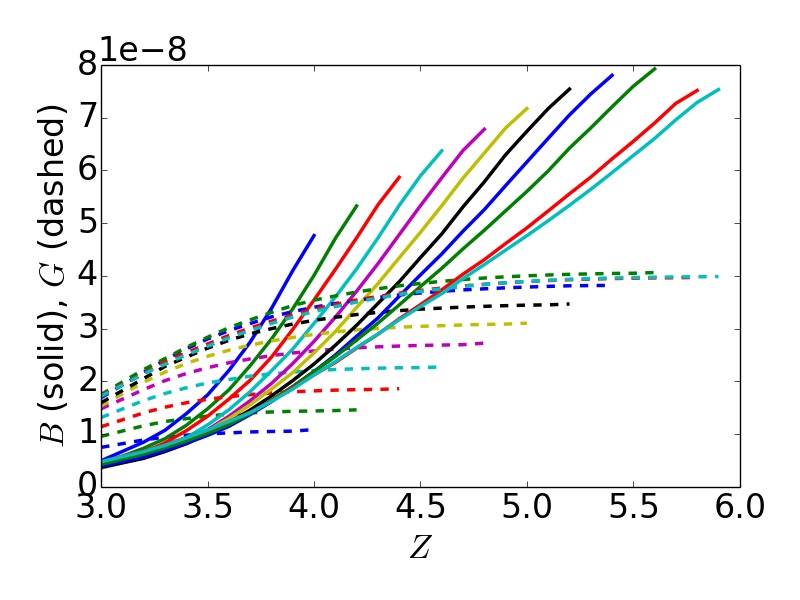}
	\caption{Bulk and shear moduli of networks as low $\Delta G$ bonds are pruned.  Initially, the bulk moduli decrease while the shear moduli remain constant, resulting in the increasing magnitude of the slope of $\nu$ observed in Fig. \ref{fig:poissons_with_prune_combo}.  After significant pruning, $G$ begins to decrease.  Near $Z=3.0$, $G/B$ plateaus and $\nu$ reaches its minimum.}
	\label{fig:GB_with_prune}
\end{figure}

\paragraph{}
To explore further how $\nu$ changes with pruning, we examine $G$ and $B$ of networks as they are pruned, as shown in Fig. \ref{fig:GB_with_prune}.  In two dimensions, linear elastic theory states $\nu=(1-G/B)/(1+G/B)$.  By pruning the lowest $\Delta G$ bonds, our aim is to maintain a high value of $G$ while reducing $B$.  During pruning, initially $B$ drops and $G$ remains nearly constant, resulting in the steepening slope of $\nu$ seen in Fig.  \ref{fig:poissons_with_prune_combo}.  At some value of $Z$ along the pruning process, $G$ begins to decrease more rapidly, and the slope of $\nu$ decreases in magnitude until $\nu$ reaches its minimum.  This accelerated decrease of $G$ can be attributed to the fact that few low $\Delta G_{i}$ bonds remain once pruning has progressed sufficiently.

\subsection*{Structural features}
Fully pruned networks ($Z=3.0$) show a range of $\nu$ values that depends on their corresponding $Z_{0}$, suggesting that there exist underlying structural differences between these pruned networks.  Figure \ref{fig:combo} shows representative networks with $Z_{0}=4.0, 4.2, 5.9$ before and after pruning to $Z=3.0$. One can appreciate that these structures are in fact quite different from each other, despite having similar numbers of nodes and bonds.  To quantify these structural differences, we calculate the percent of nodes which are ``re-entrant'' in pruned networks, as shown in Panel (b) of Fig. \ref{fig:poissons_with_prune_combo}. Here, a re-entrant node is defined as one having an angle between adjacent bonds that is greater than $180^{\circ}$.  As can be seen in Figure \ref{fig:combo}, re-entrant nodes manifest as concave angles in polygons within the network.  Such polygons tend to collapse inwards at re-entrant nodes when compressed. A sufficient number of such polygons could lead to globally auxetic behavior.  As can be seen in Fig. \ref{fig:poissons_with_prune_combo} (b), more auxetic networks exhibit a higher percentage of re-entrant nodes.  This structural motif therefore provides a basis for design of amorphous or otherwise disordered networks that are auxetic and isotropic. In this calculation we did not classify nodes with only two bonds as re-entrant, though we arrive at qualitatively the same conclusions if they are included.

\subsection*{Experimental validation}

\begin{figure}
	\centering
	\includegraphics[width=8cm]{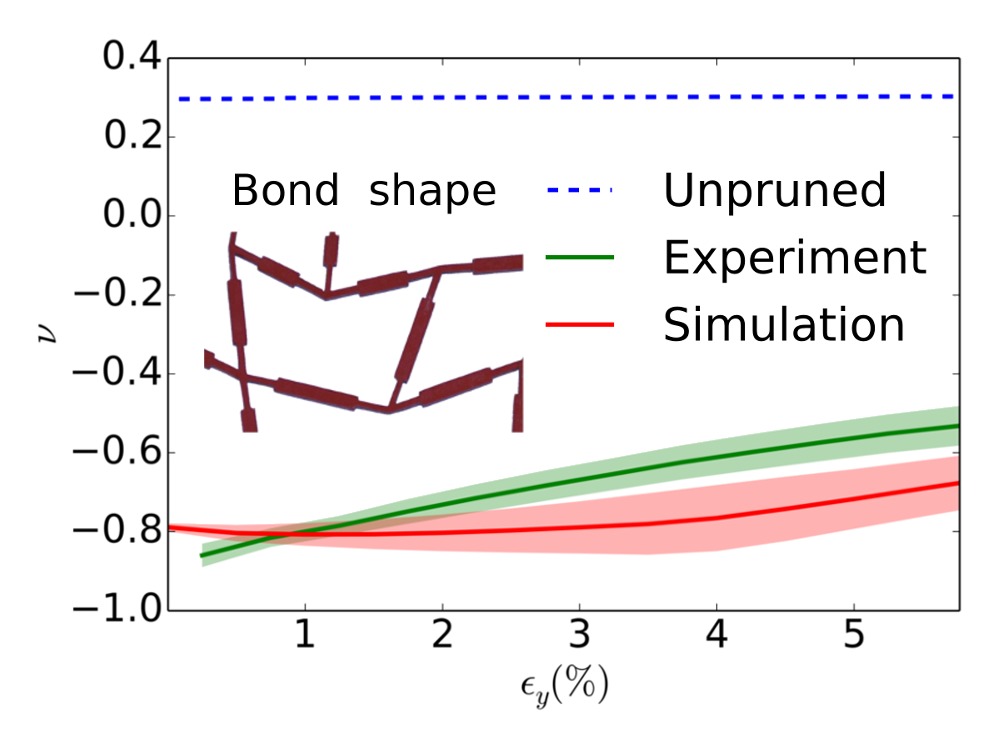}
\caption{Poisson's ratios from simulation and experiment for pruned and unpruned networks.  Shown in green and red is $\nu$ with uniaxial strain from three different networks pruned to $Z=3.0$ from $Z_{0}=5.2$.  The solid lines represent the average $\nu$ for the three configurations and the shaded areas represent standard deviations.  The dashed blue line shows $\nu$ for unpruned networks in simulation at low strain.  A value of $k_{ang}=9\e{-3}$ in simulation is fit to match this experimental bond shape, shown in the Figure.  This $k_{ang}$ fits well for all networks which use this bond shape.  A section of an experimental network is shown as an example of the individual bond shape used.}

	\label{fig:poissons_sim_exp}
\end{figure}

Experimental pruned networks are made out of laser-cut sheets of rubber\cite{rocks2017designing} as described in Methods.  The strength of bond bending, $k_{ang}$ in simulation, is modified by controlling the thickness of the bonds at the point where they attach to the nodes as well as their aspect ratio as seen in the inset to Fig. \ref{fig:poissons_sim_exp}. We focus on the bond shape shown in that figure. The deformation of such networks can be described quantitatively by our model with the value $k_{ang}=9\e{-3}$, which we use for all networks comprised of bonds of this shape.  We uniaxially compress three independent networks with $Z_{0}=5.2$ pruned to $Z=3.0$ and measure $\nu$ in both simulation and experiment as shown in Fig. \ref{fig:poissons_sim_exp}.  At low strains, networks are strongly auxetic, however $\nu$ increases monotonically with increasing strain.  Note that other experimental bond shapes are described in the Supplementary Information, including bonds which become extremely narrow near their nodes ($k_{ang}$=8\e{-3}) and bonds of constant thickness ($k_{ang}=12\e{-3}$).  With the appropriate $k_{ang}$, we find good agreement between simulation and experiment for these bond shapes as well.

\begin{figure}
	\centering
	\includegraphics[width=8cm]{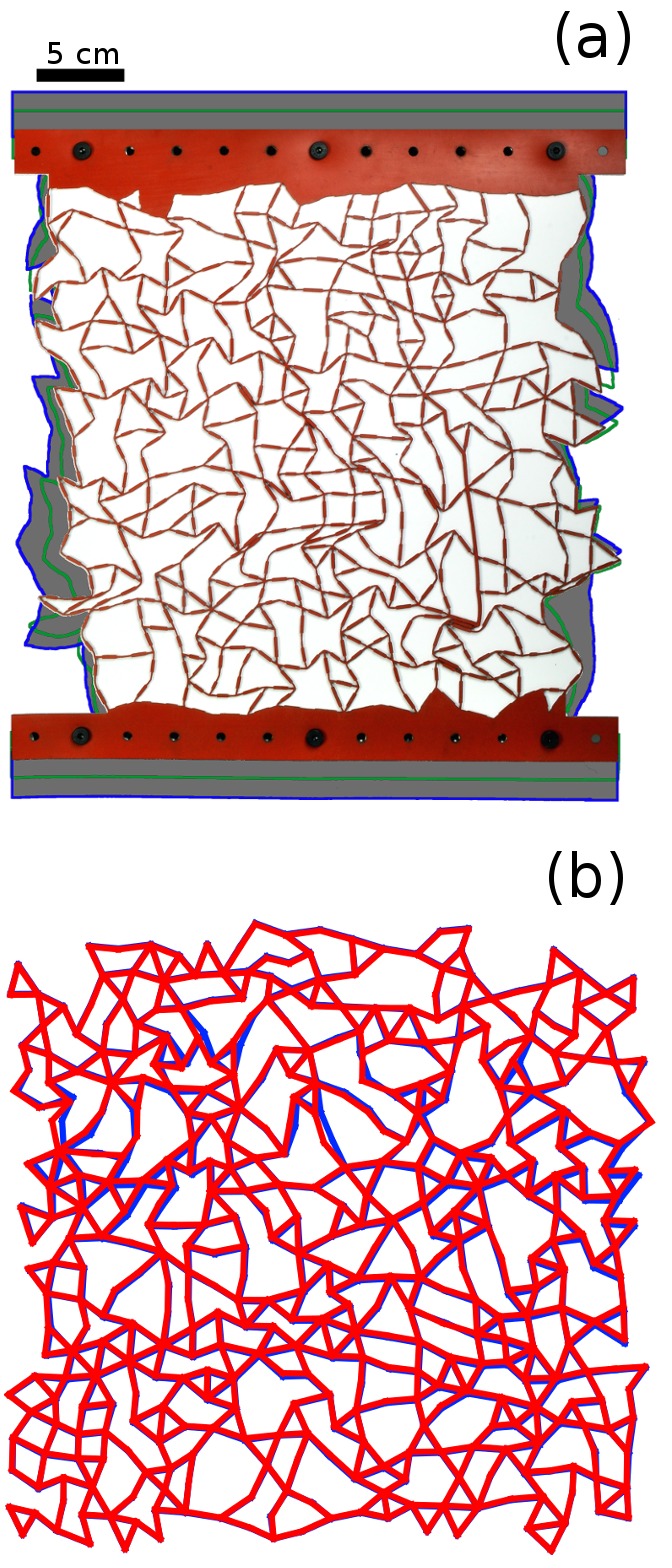}
	\caption{A compressed 500-node experimental network with comparison to simulation.  Panel (a) shows a compressed experimental network at a strain of $\epsilon_{y}=9\%$.  The gray shaded region indicates the shape of the uncompressed network, and the green outline represents the shape at $\epsilon_{y}=5\%$. Panel (b) compares an experimental configuration with that predicted from simulation at $\epsilon_{y}=2\%$.  Note that this network is isotropic and will be auxetic with respect to any uniaxial strain, which is distinct from the other networks in this work.  It shows $\nu=-0.25$ for deformations up to $\epsilon_{y}=4\%$. In red is shown a rendering of the experimental configuration and in blue is shown the simulated configuration at the same strain.  }
	\label{fig:exp_compare_config}
\end{figure}

\paragraph{}
We now examine the response of a particular network formed with $k_{ang} = 9\e{-3}$. Panel (a) of Fig. \ref{fig:exp_compare_config} shows a network compressed with $\epsilon_{y} = 0.09$.  The shape of the uncompressed network is shown in gray, serving to demonstrate its auxetic response.  Panel (b) directly compares experimental and simulated configurations at $\epsilon_{y} = 2\%$.  The experimental configuration is shown in red, and the simulated configuration is shown in blue. At this strain, experiment and simulation are in good agreement.  Note that the network pictured in Panel b is isotropic, and will be auxetic with respect to any strain.  Using pruning methods discussed in the Supplementary Information, we achieve $\nu=-0.25$ for this network.  For higher strains, our simulations are no longer able to accurately predict node positions, as they do not describe the behavior of physical bonds and nodes when they collide.  However, Despite this shortcoming, the trends of $\nu$ with $\epsilon_{y}$ are captured well by our model, as shown in Fig. \ref{fig:poissons_sim_exp}.

\subsection*{Angle bending stiffness}
\begin{figure}
	\centering
	\includegraphics[width=8cm]{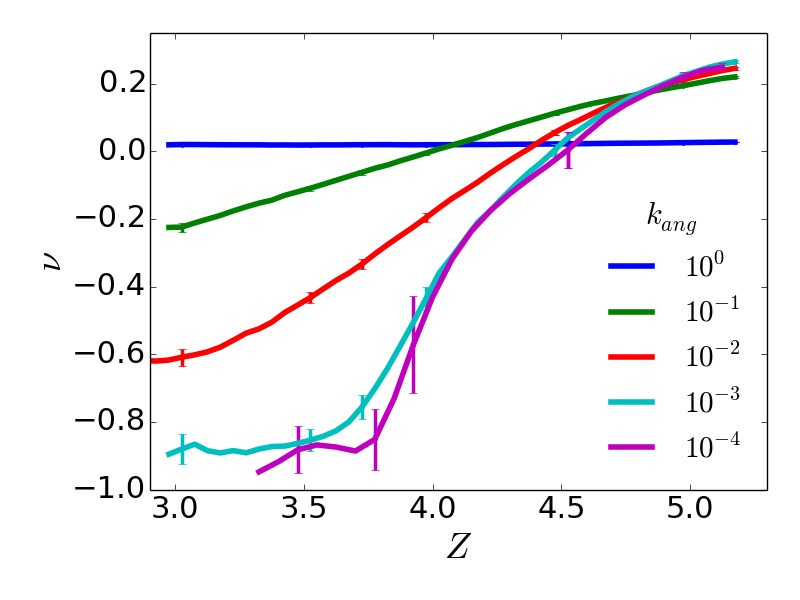}
	\caption{Poisson's ratios resulting from pruning $500$ node networks with $Z_{0}=5.2$ and $k_{ang}$ values which range from $10^{-4}$ to $10^{0}$.  With larger $k_{ang}$, pruning becomes less effective due to narrower ranges of $\Delta G_{i}$ and $\Delta B_{i}$ and increased correlations between the two quantities.  In the lower limit, $\nu$ approaches $-1.0$ as predicted in previous work\cite{goodrich2015principle}.  }
	\label{fig:sweep_kang}
\end{figure}

We have focused only on values of $k_{ang}$ within a relatively narrow range, but one could conceive of specially designed experimental realizations which would span a much wider range.  This is of interest because networks with greater bond stiffness can withstand greater strains before failing as shown in Supplementary Information. As such, we turn our attention to the effect that a wide range of bond bending stiffness has on $\nu$.  We study networks with $Z_{0}=5.2$, which yielded the lowest value of $\nu$ for $k_{ang}=0.01$.  Figure \ref{fig:sweep_kang} shows $\nu$ resulting from low $\Delta G$ pruning of $500$-node networks with values of $k_{ang}$ that span five orders of magnitude, from $10^{-4}$ to $10^{0}$.  Consistent with previous work \cite{goodrich2015principle}, $\nu \rightarrow -1$ as $k_{ang} \rightarrow 0$ in fully pruned networks. Consistent with these results, we find that as $k_{ang}$ becomes smaller, bond response distributions become wider and are more easily modified by pruning, as shown in Supplementary Information Fig 1.  The value of $\nu$ in pruned networks changes smoothly from $-1$ to roughly zero as $k_{ang}$ approaches the coefficient for bond compression.  To understand this transition, we must as before examine both the distributions and correlations of $\Delta G_{i}$ and $\Delta B_{i}$.    As shown in Supplementary Information, materials with  $k_{ang}>1$ are also auxetic because the strong angles preserve the networks' shape.   

\subsection*{Stress-strain behavior}

For a variety of impact-mitigation applications, it is of interest to develop materials that display a relatively constant stress-strain behavior. Such materials can absorb more energy while maintaining lower applied forces, and thus reduce the possibility of damage.  As shown in Supplementary Information Fig 3, pruned networks display nearly constant stress past $3\%$ strain.  At such strains, linear response calculations are no longer accurate, as shown in Supplementary Information Fig 2.  We find that the linear response framework applies well until roughly $1\%$ strain.

\subsection*{Bond strength optimization}
\begin{figure}
	\centering
	\includegraphics[width=8cm]{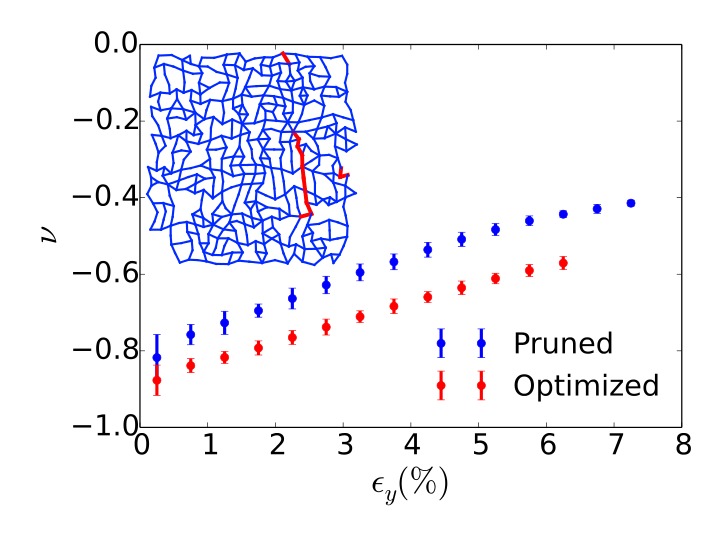}
	\caption{Experimentally measured $\nu$ for an ordinary pruned network and an otherwise identical network in which several of the bonds have been strengthened.  Error bars show $2.5 \sigma$. The bonds which were strengthened were chosen in simulation in order to reduce $\nu$.  In simulation, the networks with regular and strengthened bonds were predicted to show $\nu=0.79, 0.91$ respectively at small strains.  The inset network configuration shows in red, bonds which were strengthened in experiment and in blue, unstrengthened bonds}
	\label{fig:poissons_reg_strength}
\end{figure}

Up to this point, we have relied on homogeneous materials, with identical bonds, for all calculations and experiments. In what follows, we modify the strength of individual bonds as a means for decreasing $\nu$ in networks composed of bonds with different stiffnesses.  This process can be mimicked in experiment by modifying the thickness or material of a given bond.  We implement a simple optimization algorithm that iteratively strengthens the bond leading to the greatest decrease in $\nu$, as described in the Supplementary Information.  Both the compressive and bending modulus of a particular bond are increased when a bond is strengthened. We examine a particular network with $k_{ang}=10^{-2}$ and $\nu=-0.79$.  By successively strengthening individual bonds in this network, we further decrease $\nu$ from $-0.79$ to $-0.91$ in simulation.  Interestingly, after $430$ iterations with $649$ total bonds in the network (where one bond in strengthened by 10\% at each iteration), $94\%$ of the bonds remained untouched, while a select few, $1.8\%$, are strengthened to more than five times their original strength, leading to an essentially bimodal distribution of bond strengths. These strengthened bonds are almost all connected as shown in the inset of Fig. \ref{fig:poissons_reg_strength}.

\paragraph{}
To validate the predictions of our simulations, we also prepared an experimental realization of this optimized network.  For simplicity, bonds strengthened by a factor of five or greater were made thicker, and others were left unchanged. The corresponding experimental values of $\nu$ are shown in Fig. \ref{fig:poissons_reg_strength}, showing a decrease in $\nu$ of $0.059$ at $\epsilon_{y}=0.25\%$ and a decrease of $0.11$ at $\epsilon_{y}=2.75\%$, in good agreement with predictions. Importantly,
these optimized materials with a few significantly stronger bonds lend themselves to additive manufacturing.  In such realizations, some bonds could be constructed of highly rigid materials, while the remainder would be more pliable, and more advanced optimization algorithms could readily be applied to this problem.
	
\showmatmethods{} 
\section*{Conclusion}
In summary, we have establish that it is possible to create designer auxetic materials from amorphous networks. The models and concepts introduced in this work have been validated through a concerted program of design, computation and laboratory experimentation. Amorphous networks are shown to offer a number of control parameters that can be tuned to achieve particular mechanical responses. It is found, for example, that a networks' propensity to be made auxetic depends on both the network's original coordination number, as well as the relative resistance to angle bending. More pliable networks yield the lowest Poisson's ratios due to their wide bond response distribution and their low response correlation. Stiffer networks are less amenable to pruning, and only show limited changes of their Poisson's ratio through pruning. By relying on bond-strength optimization schemes, however, it is possible to alter the Poisson's ratio of networks with stiff bonds considerably, thereby providing a strategy to alter not only how auxetic a material is, but also its intrinsic stiffness. While the results presented here have been limited to two-dimensional networks, the concepts and strategies proposed should be equally applicable to three dimensions. We therefore anticipate that they could be potentially useful for applications involving additive manufacturing.

\section*{Methods}
Simulated networks are generated as described in the Models section.  A harmonic wall coefficient of $2.0$ is used to compress particles.  To measure $\nu$, $\epsilon_{y} = 1\e{-4}$ is applied and the transverse strain of nodes at the left and right edges of the network is measured.  Bulk properties are measured by applying uniform compressions of $1\e{-4}$.  Shear properties are measured with $\epsilon_{x}=-1\e{-4}$ and $\epsilon_{y}=1\e{-4}$ or $\gamma=1\e{-4}$, for pure and simple shear, respectively.  The average force is relaxed to $1\e{-13}$ for all measurements.  To mimic experiment, particles along the top and bottom edges of networks are restrained in the $x$ dimension.
The coefficient to describe bond bending, $k_{ang}$ is fit by determining the value of $k_{ang}$ which minimized mean square distance between nodes in between uniaxially strained experimental and simulated networks at $\epsilon_{y}=3\%$.  The same $k_{ang}$ value is used to describe each class of experimental bonds.

\subsection*{Experimental Methods}
Experimental networks are constructed out of laser-cut silicone rubber sheets with a Shore value of $A 70$ and a thickness of $1.5$ mm as described in previous work \cite{rocks2017designing}. We can vary the relative resistance to angle bending (which is quantified by $k_{ang}$ in our simulation model), by narrowing or widening a section of the bond near the node (see Fig. \ref{fig:exp_compare_config}).  To facilitate measurement, nodes at the top and bottom of the network are fused into a solid rubber piece, as shown in Panel (a) of Fig. \ref{fig:exp_compare_config}.  The Poisson's ratio is determined by applying a uniaxial compression in the $y$ direction and measuring the resulting lateral strain.

\acknow{We gratefully acknowledge Carl Goodrich and Daniel Hexner for their helpful discussions.  The design and fabrication of mechanical metamaterials based on random networks was supported by the University of Chicago Materials Research Science and Engineering Center, which is funded by the National Science Foundation under award number DMR-1420709. The development of auxetic systems for impact mitigation applications and the corresponding materials optimization strategies presented here are supported by the Center for Hierarchical Materials Design (CHiMaD).}

\showacknow{}

\bibliography{pnas-sample}

\pagebreak

\section*{Supplementary Information}

\subsection*{Bond response distributions with bending stiffness}
\begin{figure}[h]
	\centering
	\includegraphics[width=8cm]{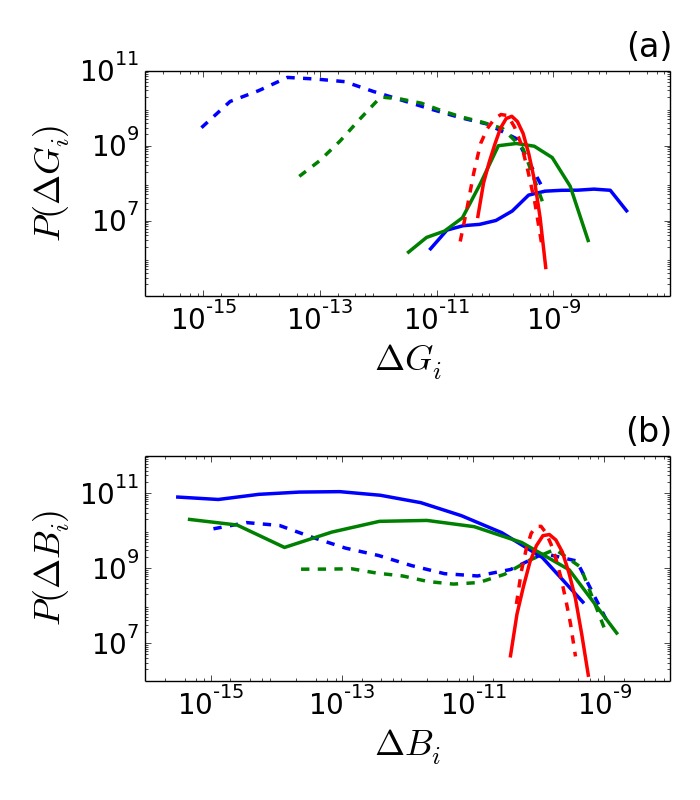}
	\caption{Distributions of $\Delta G_{i}$, $\Delta B_{i}$ for networks before pruning ($Z=5.2$) and after pruning ($Z=3.5$) of networks with different values of $k_{ang}$.  Distributions are shown for $k_{ang}=$ $10^{-4}$ (blue), $10^{-2}$ (green), $10^{0}$ (red).  Distributions for unpruned networks are shown with dashed lines, while those for pruned networks are shown with solid lines.  As $k_{ang}$ decreases, the change in distributions becomes more significant, corresponding to the greater change in $\nu$ with pruning as seen in Figure 8 of the main text. }
	\label{fig:stressDistsPrune}
\end{figure}

The effect of angle-bending stiffness can be appreciated by examining the bond response distributions when $k_{ang}$ is varied from $10^{-4}$ to $10^{0}$, as shown in Figure \ref{fig:stressDistsPrune}.  Dashed lines denote the distributions of unpruned networks with $Z_{0}=5.2$.  As angles become stiffer, the distribution of bond responses narrows significantly.  As mentioned in the main text, broader distributions enable more effective pruning.  The solid lines of Figure \ref{fig:stressDistsPrune} show the bond response distributions for the same networks after pruning to $Z=3.5$.  Not only do stiffer bonds lead to narrower distributions - they also make these distributions more difficult to modify through pruning. For the stiffest angles considered here, $k_{ang}=10^{0}$, no significant change is observed in the corresponding distributions, and little change occurs in $\nu$.  A angles become easier to deform (more compliant), the change in distributions becomes more dramatic. Correlations between bond responses $\Delta G_{i}$ and $\Delta B_{i}$ decrease significantly, leading to larger changes in $\nu$. We calculate the Pearson correlation coefficient for unpruned networks to be $0.134$, $0.125$, and $0.843$ for $k_{ang}=$ $10^{-4}$, $10^{-2}$, and $10^{0}$, respectively.  The lower the correlation, the more tunable the network becomes. 

\subsection*{Non-linear behavior}

\begin{figure}
	\centering
	\includegraphics[width=8cm]{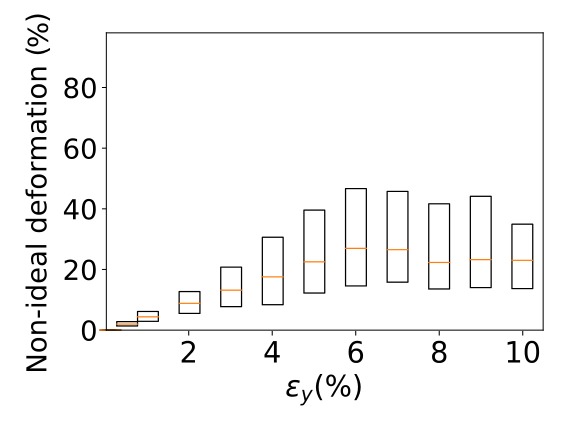}
	\caption{Relative error of between node displacements as predicted by molecular dynamics and linear regime calculations with uniaxial strain.  Relative error is calculated as $100 * \lvert \vec{r}_{md} - \vec{r}_{lin} \lvert/\lvert \vec{r}_{md} \lvert$.  Red lines denote the medians of samples, box edges denote the $25^{th}$ and $75^{th}$ percentile.}
	\label{fig:linear}
\end{figure}

While a host of properties are readily accessible through linear-response calculations \cite{goodrich2015principle,sussman2016spatial}, the corresponding predictions are not valid for large deformations.  To address the extent of linear response, we compare in Figure \ref{fig:linear} the node positions of systems strained along the $y$-direction - as predicted by linear-regime calculations - to those predicted by the true network-dynamics formalism adopted here.  We quantify the relative error between the two predictions through the average of $100 * \lvert \vec{r}_{md}-\vec{r}_{lin} \lvert/\lvert \vec{r}_{md} \lvert$, where $\vec{r}_{md}$ is the calculated displacement of a node and $\vec{r}_{lin}$ is 
the displacement of a node within the linear regime scaled to the strain of interest.  The linear-regime responses are calculated at $\epsilon_{y}=0.1\%$. We average this quantity across every node in three distinct $500$-node networks with $Z=3.0$ pruned from $Z_{0}=5.2$. The relative error increases nearly linearly from roughly $0\%$ at $\epsilon_{y}=0.5 \%$ to a median of $30\%$ at $\epsilon_{y}=5\%$. Note, however, that the relative error of individual nodes can exceed $800\%$.  From this data, one can assess to what degree linear regime predictions can be considered accurate over a range of strains.

\subsection*{Stress-strain behavior}

\begin{figure}
	\centering
	\includegraphics[width=8cm]{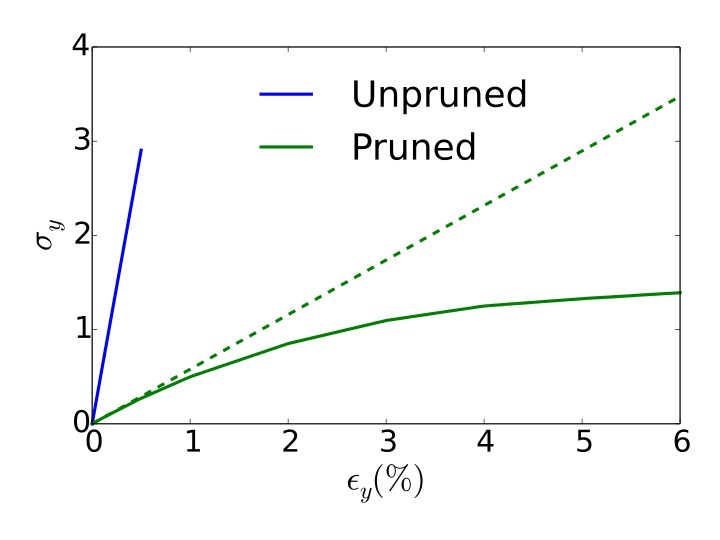}
	\caption{Stress-strain curve for unpruned ($Z=5.2$) and pruned ($Z=3.0$)  networks.  The dashed green line shows an extrapolation of the pruned networks' linear regime behavior. Error bars are smaller than the size of the line.  The unpruned network stress behaves linearly with strain, as is expected from a network which employs harmonic constraints.  The pruned network, however, behaves sublinearly, as is characteristic of a material undergoing complex rearrangements.}
	\label{fig:stress_strain}
\end{figure}

 For a variety of impact-mitigation applications, it is of interest to develop materials that display a relatively constant stress-strain behavior. Such materials can absorb more energy while transmitting less force, and thus reduce the possibility of damage.  Figure \ref{fig:stress_strain} shows the stress-strain behavior of networks before and after pruning.  Unpruned networks show typical linear stress versus strain behavior.  Pruned networks, however, behave sub-linearly, and exhibit a nearly constant stress past 3$\%$ strain.  This behavior is characteristic of a material undergoing complex rearrangements, such as collapsing concave structures.  Also shown is an extrapolation of the linear-regime behavior for pruned networks, demonstrating the degree of sublinearity.
 
 \subsection*{Bond strength optimization algorithm}

To further reduce $\nu$ of pruned networks, we employ a simple optimization algorithm that works as follows:  At each iteration, we measure the change in $\nu$ resulting from strengthening each bond by $10\%$.  We then strengthen the bond which leads to the greatest decrease in $\nu$ by $10\%$ and repeat the algorithm.  Both the compressive and angular component of the bond are strengthened.  The results of this optimization are shown in the main text.
 
 \subsection*{Isotropic and anisotropic networks}
 
\begin{figure*}[t!]
 	\centering
   	\includegraphics[width=12cm]{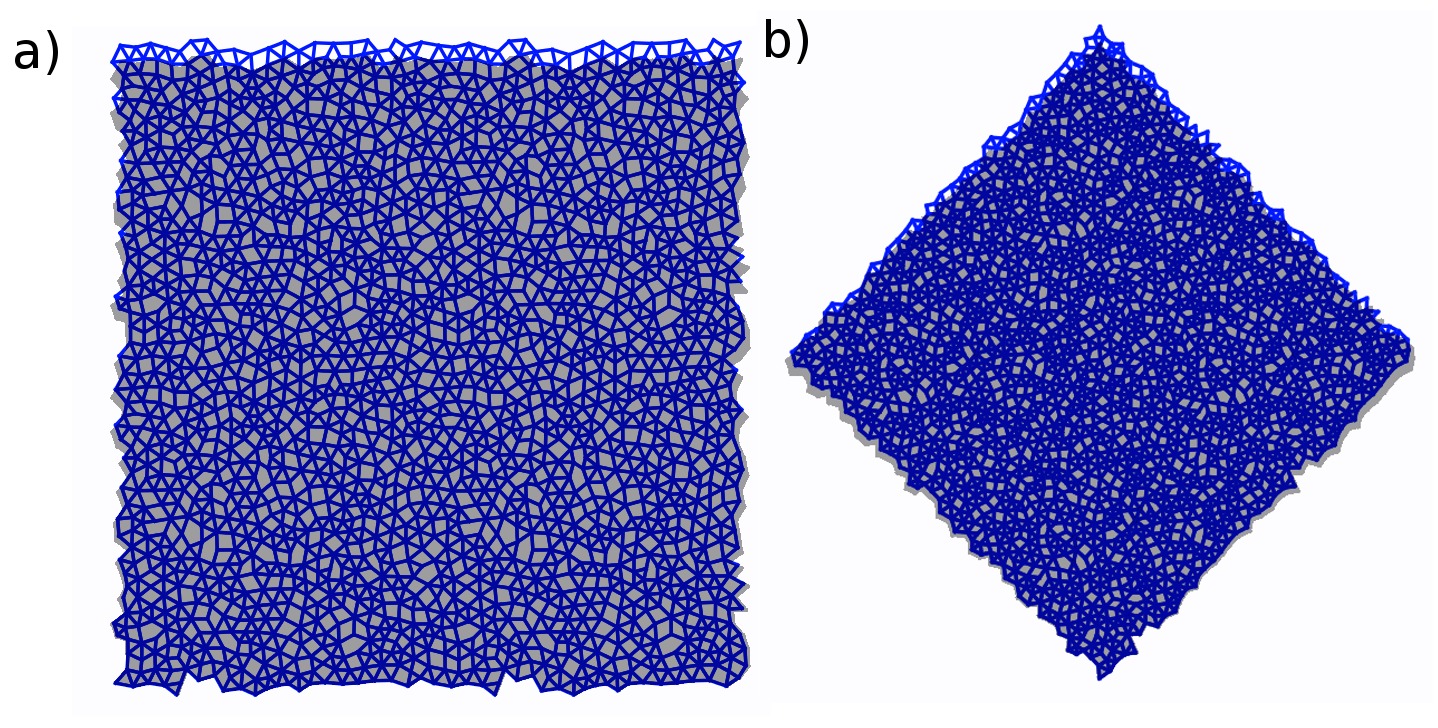}
	\caption{Unpruned networks deformed by pulling along the vertical axis.  The gray shaded region shows the shape of the undeformed network. This network shows positive $\nu$ with respect to both deformations, since it is unpruned.}
	\label{fig:unpruned}
\end{figure*}

\begin{figure*}[t!]
 	\centering
   	\includegraphics[width=12cm]{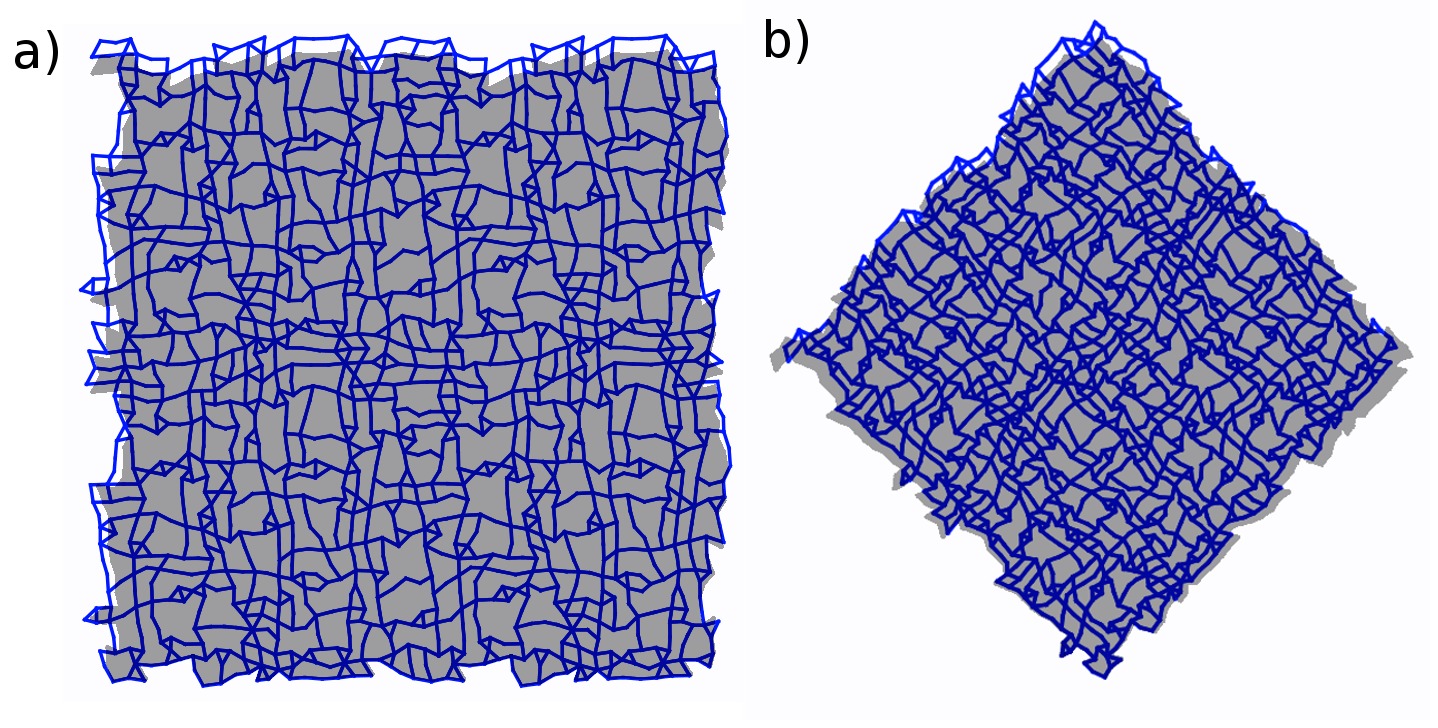}
	\caption{Low $\Delta G_{p}$ pruned network deformed by pulling along the vertical axis.  The gray shaded region shows the shape of the undeformed network. This network is auxetic with respect to deformation normal to its top and bottom edge, but not when pulled from its corners. }
	\label{fig:pure}
\end{figure*}

\begin{figure*}[t!]
 	\centering
   	\includegraphics[width=12cm]{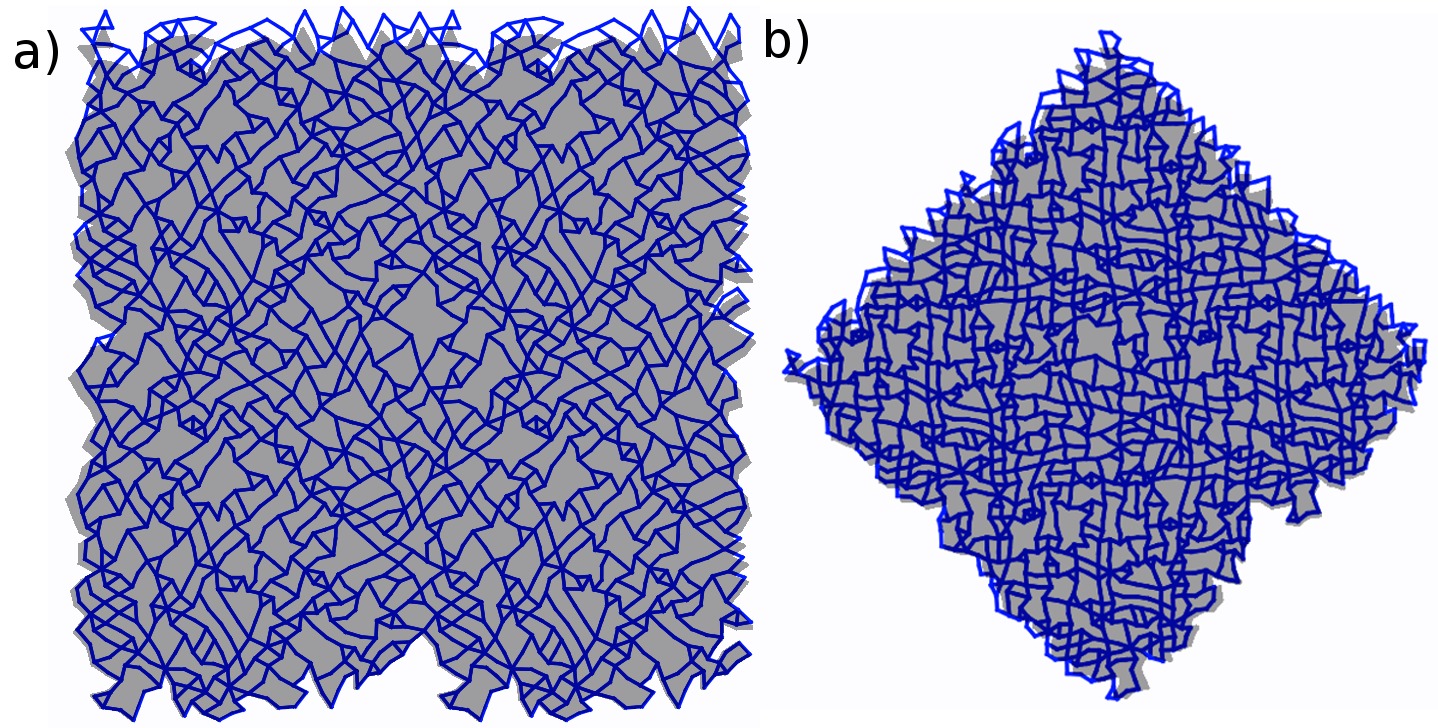}
	\caption{Low $\Delta G_{s}$ pruned network deformed by pulling along the vertical axis.  The gray shaded region shows the shape of the undeformed network.  This network is auxetic with respect to being pulled from its corners, but not deformation normal to its top and bottom edge. }
	\label{fig:simple}
\end{figure*}

 In the main text, we discuss the two relevant shear moduli - those associated with pure and simple shear.  We denote these moduli $G_{p}$ and $G_{s}$, respectively. Here we show that networks pruned only to have high $G_{p}/B$ do not have high $G_{s}/B$, and vice versa.  In practice, this means that a network pruned to be auxetic when pulled in a direction normal to its edges will not be auxetic if pulled outwards from its corners.  We also examine isotropic networks - those which have been pruned to show low $(G_{p} + G_{s})/B$.  Isotropic networks are formed by iteratively pruning the minimum $\Delta G_{p} + \Delta G_{s}$ bond.
 \paragraph{}
 In what follows, we examine what happens when unpruned, low $\Delta G_{p}$ pruned, low $\Delta G_{s}$ pruned and low $\Delta G_{p} + \Delta_{s}$ pruned (isotropic) networks are deformed by stretching the materials in a direction normal to their edges and by stretching along a diagonal. Figure \ref {fig:unpruned} shows an unpruned network deformed by pulling from the edges and corners (along a diagonal) in Panels a and b, respectively.  The gray background shows the outline of the undeformed network.  Since the network is unpruned, $\nu$ is positive when pulled from the top and bottom edges and from the corners.  Figure \ref{fig:pure} shows a network which has been low $\Delta G_{p}$ pruned.  As can be seen, the network is auxetic when pulled from the top and bottom edges as in Panel a, but not when pulled from the corners, as in Panel b.  If we instead prune low $\Delta G_{s}$ bonds as in Figure \ref{fig:simple}, we observe the opposite behavior.  The network is not auxetic when pulled from the top and bottom edges as in Panel a, and it is auxetic when pulled from the corners as in Panel b.  
 \paragraph{}
 
\begin{figure*}[t!]
 	\centering
   	\includegraphics[width=12cm]{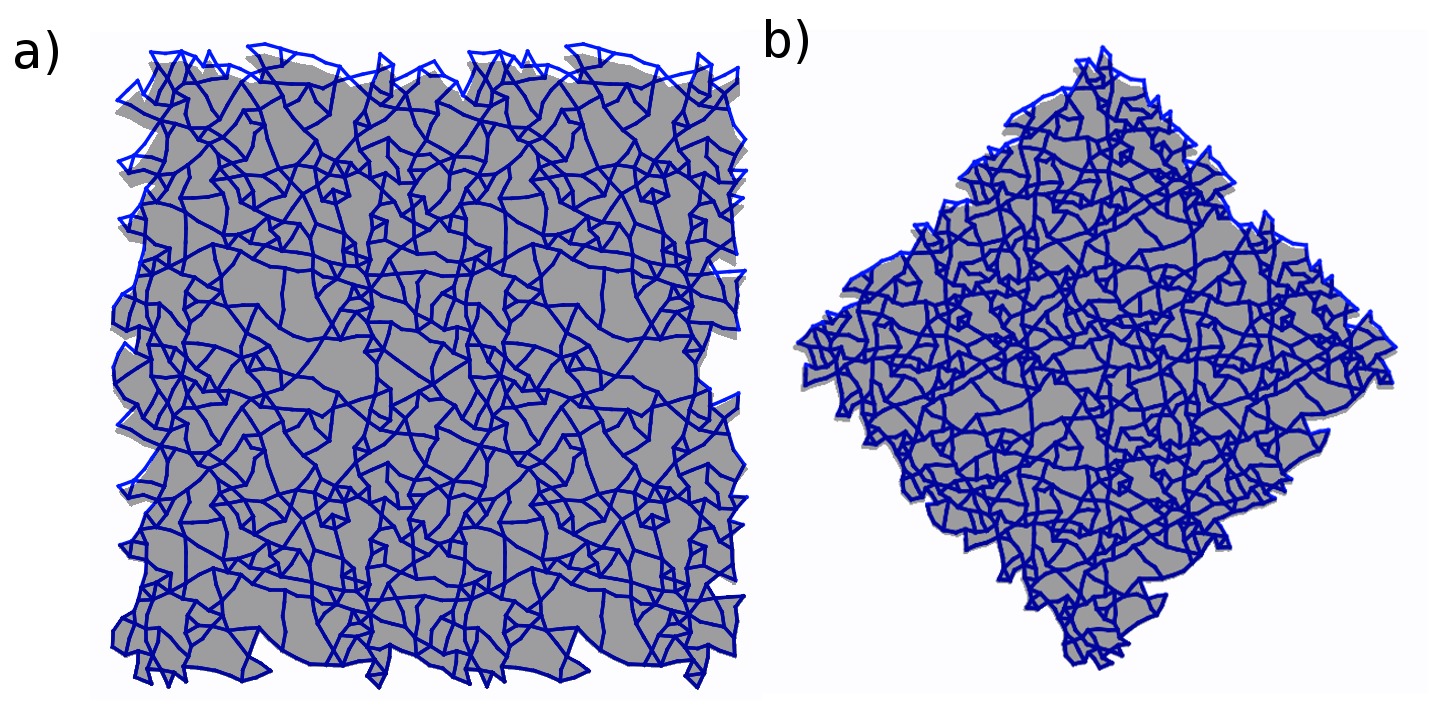}
	\caption{Low $(\Delta G_{s}+\Delta G_{p})$ pruned network deformed by pulling along the vertical axis.  The gray shaded region shows the shape of the undeformed network.  This network is isotropic, so it auxetic for all types of deformation. However $\nu$ of this network is higher than in anisotropic networks.}
	\label{fig:iso}
\end{figure*}

 Figure \ref{fig:iso} shows an isotropic network formed by iteratively pruning the lowest $\Delta G_{p} + \Delta G_{s}$ bond.   Creating isotropic networks is a more demanding optimization task, since two moduli must be kept high relative to $B$ instead of only one.  As a result, $\nu$ of the isotropic networks is less negative than $\nu$ of anisotropic networks.  The network shown is predicted to have $\nu=-0.25$ isotropically in the linear regime using $k_{ang}=0.01$ as in other networks.  The structure of this isotropic network is noticeably different from that of anisotropic networks - while anisotropic networks show significant rectangular ordering, the isotropic sample does not.


 \subsection*{Effect of angle potentials on $\nu$}
\begin{figure}
	\centering
	\includegraphics[width=8cm]{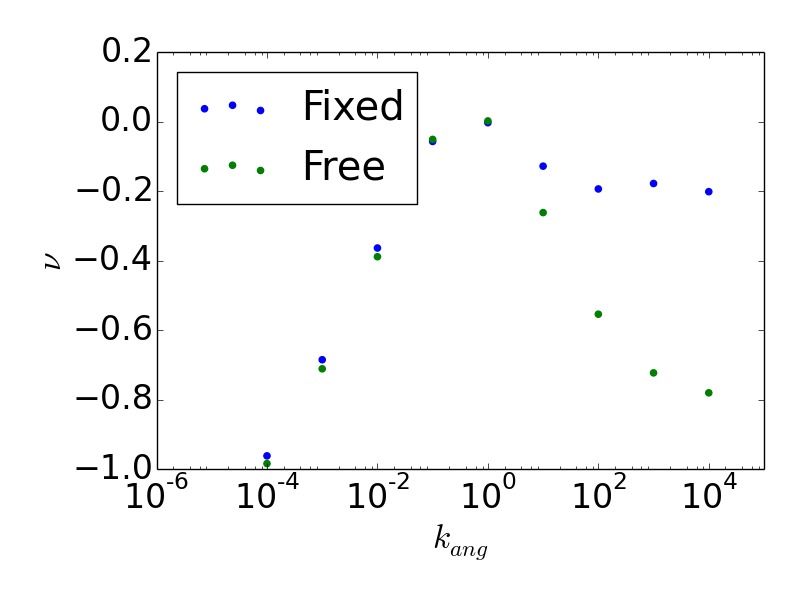}
	\caption{Poisson's ratio of a single network pruned to $Z=3.87$ by low $\Delta G$ pruning at $k_{ang}=10^{-4}$.  The value of $k_{ang}$ is then swept from $10^{-4}$ to $10^{4}$ with both fixed and free boundary conditions.  }
	\label{fig:sweep_kang_single}
\end{figure}

Initial attempts to create auxetic networks in experiment were based on simulations that did not include angle-bending forces. In this work, we incorporate the effect of angle-bending forces as described in the main text.  Here, we investigate what occurs when $k_{ang}$ is varied for already-pruned networks.  The results are shown in Figure \ref{fig:sweep_kang_single}. We start with a single $500$ node network pruned to $Z=3.87$ with $k_{ang}=10^{-4}$ (analogous to simulations that ignored $k_{ang}$).  As we increase $k_{ang}$, $\nu$ follows the same trends as networks pruned with those values of $k_{ang}$, as shown in Fig. 8 of the main text.    
\paragraph{}
At $k_{ang}=1$, the coefficients for angle bending and compression are equal.  Interestingly, $\nu$ passes through zero at this point.  As angles-bending forces becomes even stronger, the network retains its shape as deformed, leading to $\nu<0$. 
\paragraph{}
Therefore there are two different mechanisms that can produce auxetic behavior in these networks.  At low $k_{ang}$, networks can freely deform allowing for concave polygons to collapse, yielding $\nu<0$.  As high $k_{ang}$ rigid angles maintain network shape as it is strained.  This also yields $\nu<0$.  At the point where the networks `switch' between these two competing mechanisms, $\nu=0$.

 \subsection*{Boundary conditions}
\begin{figure}
	\centering
	\includegraphics[width=8cm]{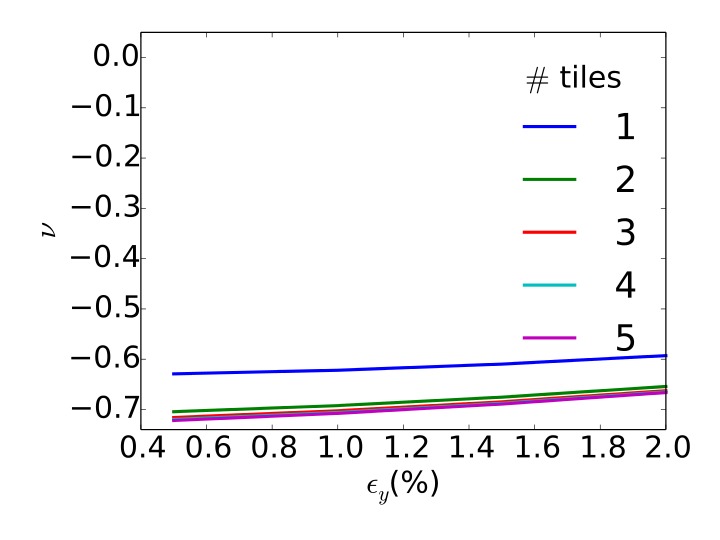}
	\caption{Poisson's ratio vs. uniaxial strain of simulated 500 node periodic networks which have been tiled in both dimensions and then made finite.  Increased tiling decreases $\nu$, as edge effects are reduced.  These edge effects are not present when the periodic system is pruned.  For systems of this  size $\nu$ rapidly converges to a final value with tiling.}
	\label{fig:nu_with_tile}
\end{figure}

\begin{figure}
	\centering
	\includegraphics[width=8cm]{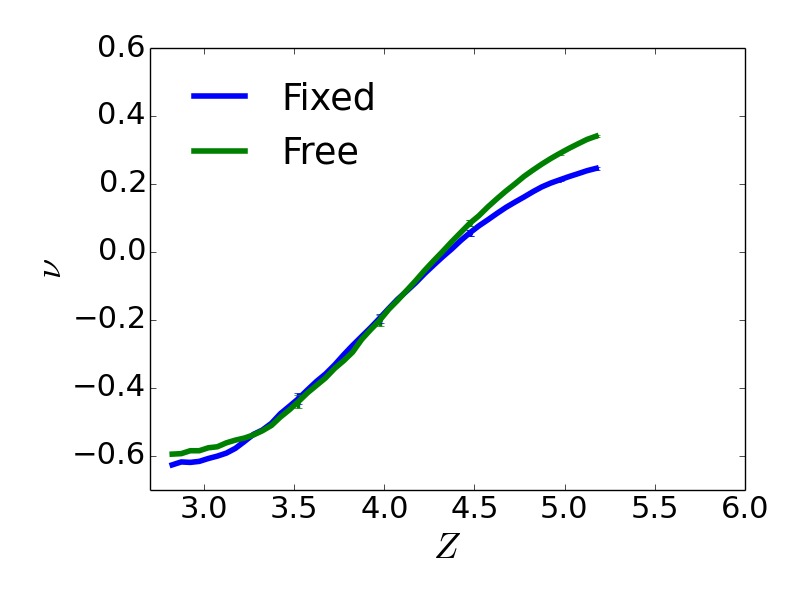}
	\caption{Poisson's ratio of networks pruned with fixed and free boundary conditions for $500$ node networks.  The boundary conditions do not significantly effect the resulting values of $\nu$.  }
	\label{fig:fixed_free_bounds}
\end{figure}

The boundary conditions of the networks in this study differ in two ways from previous attempts.  In previous attempts, the simulations were periodic.  In experimental realizations, however, there are no periodic boundaries and these periodic systems were terminated, with their top and bottom nodes fixed in the $x$ dimension.  In this work, we incorporate these experimental features into the simulation in order to match the experiments more closely.  Figure \ref{fig:nu_with_tile} shows $\nu$ vs. $\epsilon_{y}$ for a periodic $500$ node network which has been cast as a finite system tiled in both dimensions some number of times.  The Poisson's ratio quickly reaches an asymptotic value as the tiling increases.  However if a periodic system is simply made finite with no tiling (as is the case for $\#tiles = 1$), $\nu$ will be non-trivially less negative.  
\paragraph{}
Figure \ref{fig:fixed_free_bounds} shows $\nu$ of networks with fixed and free boundary conditions pruned with low $\Delta G$ pruning from $Z=5.2$ to $Z=3.0$.  In networks with fixed boundary conditions, nodes at the top and bottom of the network are restrained in both the $x$ and $y$ dimension upon deformation, mimicking experimental boundary conditions.  In networks with free boundary conditions, nodes at the top and bottom of the network may relax in the $x$ dimension upon deformation. As can be appreciated, the fixed boundary conditions do not significantly affect $\nu$. All results in this work are produced using fixed boundary conditions unless otherwise specified.

\end{document}